# Stretched or noded orbital densities and self-interaction correction in density functional theory

Chandra Shahi[1], Puskar Bhattarai[1], Kamal Wagle[1], Biswajit Santra[1], Sebastian Schwalbe[2], Torsten Hahn[2], Jens Kortus[2], Koblar A. Jackson[3], Juan E. Peralta[3], Kai Trepte[3], Susi Lehtola[4], Niraj K. Nepal[1], Hemanadhan Myneni[1], Bimal Neupane[1], Santosh Adhikari[1], Adrienn Ruzsinszky[1], Yoh Yamamoto[5], Tunna Baruah[5], Rajendra R. Zope[5], and John P. Perdew[1,6]

[1]Department of Physics, Temple University, Philadelphia, PA 19122 USA

[2]Institute of Theoretical Physics, TU Bergakademie Freiberg, Leipziger Str. 23, D-09596, Freiberg, Germany

[3]Department of Physics and Science of Advanced Materials, Central Michigan University, Mount Pleasant, MI 48859 USA

[4]Department of Chemistry, FI-00014 University of Helsinki, Finland

[5]Department of Physics, University of Texas at El Paso, El Paso, TX 79968 USA

[6]Department of Chemistry, Temple University, Philadelphia, PA 19122 USA

Abstract: Semi-local approximations to the density functional for the exchange-correlation energy of a many-electron system necessarily fail for lobed one-electron densities, including not only the familiar stretched densities but also the less familiar but closely-related noded ones. The Perdew-Zunger (PZ) self-interaction correction (SIC) to a semi-local approximation makes that approximation exact for all one-electron ground- or excited-state densities and accurate for stretched bonds. When the minimization of the PZ total energy is made over real localized orbitals, the orbital densities can be noded, leading to energy errors in many-electron systems. Minimization over complex localized orbitals yields nodeless orbital densities, which reduce but typically do not eliminate the SIC errors of atomization energies. Other errors of PZ SIC remain, attributable to the loss of the exact constraints and appropriate norms that the semi-local approximations satisfy, and suggesting the need for a generalized SIC. These conclusions are supported by calculations for one-electron densities, and for many-electron molecules. While PZ SIC raises and improves the energy barriers of standard generalized gradient approximations (GGA's) and meta-GGA's, it reduces and often worsens the atomization energies of molecules. Thus PZ SIC raises the energy more as the nodality of the valence localized orbitals increases from atoms to molecules to transition states. PZ SIC is applied here in particular to the SCAN meta-GGA, for which the correlation part is already self-interaction-free. That property makes SCAN a natural first candidate for a generalized SIC.

1. **Introduction: semi-local approximations and self-interaction correction**

Kohn-Sham density functional theory [1,2] simplifies the many-electron ground-state problem of condensed-matter physics or quantum chemistry to a self-consistent one-electron form, in a way that is formally exact for the total energy and electron density. To make this theory tractable and widely useful, an approximation (typically a computationally-efficient semi-local one) must be made to the exact density functional for the exchange-correlation energy, the many-body "glue" that binds atoms to form molecules and solids.

A semi-local approximation expresses this energy as a single integral over three-dimensional space of a function of various ingredients that are available from the solution of the self-consistent one-electron Schrödinger equation. The local spin density approximation (LSDA) [1,3] uses only the local electron spin densities as ingredients, the generalized gradient approximation (GGA) adds their gradients, and the meta-GGA further adds the positive spin-resolved orbital kinetic energy densities:

$$E_{xc}^{approx}[n_\uparrow, n_\downarrow] = \int dr\, n\varepsilon_{xc}^{approx}(n_\uparrow, n_\downarrow, \nabla n_\uparrow, \nabla n_\downarrow, \tau_\uparrow, \tau_\downarrow). \quad (1)$$

Square brackets in Eq. (1) indicate a functional, and round brackets a function. The total electron density $n(r) = n_\uparrow(r) + n_\downarrow(r)$ is the sum of up- and down-spin contributions.

The integrand of Eq. (1) can be constructed empirically by fitting to molecular data, or in a more first-principles way by satisfying known exact constraints and appropriate norms. Within a semi-local form, many but not all exact constraints or mathematical properties of the exact functional can be satisfied for all possible electron densities by the careful design of the integrand. The uniform electron gas [3] is an appropriate norm for the LSDA, for the Perdew-Burke-Ernzerhof (PBE) GGA [4], and for the strongly constrained and appropriately normed (SCAN) meta-GGA [5] that will be considered here. Free atoms (but not molecules) provide other appropriate norms for SCAN, which further satisfies all 17 known exact constraints that a semi-local form can satisfy. All three ingredients $n_\sigma, |\nabla n_\sigma|, \tau_\sigma$ in Eq. (1) are needed to recognize one- and two-electron regions for which there are special constraints (e.g., the correlation energy vanishes for any one-electron density). However, once a meta-GGA has been constructed, it is possible if desired to approximate $\tau_\sigma$ in terms of $n_\sigma, |\nabla n_\sigma|, \nabla^2 n_\sigma$ [6].

Recent successes, e.g., Refs. [7-12], suggest that SCAN may be approaching the limiting accuracy of semi-local functionals. It may be time to satisfy more exact constraints via fully-nonlocal functionals. Semi-local spin-density functionals cannot be exact for all one-electron (fully spin-polarized) densities $n$, for which the exact spin-density functional for the exchange-correlation energy must exactly cancel the fully nonlocal Hartree electrostatic energy

$$U[n] = \frac{1}{2}\int dr\, n(r)\int dr'\, n(r')/|r' - r|. \quad (2)$$

For a one-electron density, all of the $U[n]$ term in the total energy is a spurious self-interaction. The residual self-interaction error of a semi-local functional for a one-electron density, due to an imperfect cancellation of $U[n]$ by the semi-local exchange-correlation energy, manifests in serious errors for stretched bonds that can arise in binding energy curves (e.g., for $H_2^+$ [13,14], as shown in Fig. 1]), charge transfers [15], energies of transition states for chemical reactions [16], and in other ways.

In 1981, Perdew and Zunger [17] proposed a self-interaction correction to any spin-density functional approximation:

$$E_{xc}^{approx-SIC} = E_{xc}^{approx}[n_\uparrow, n_\downarrow] - \sum_{i\sigma}^{occup}\{U[n_{i\sigma}] + E_{xc}^{approx}[n_{i\sigma}, 0]\}. \qquad (3)$$

Here $n_{i\sigma}(\mathbf{r})$ is the $i$-th occupied orbital density of spin σ, and thus a one-electron density. The resulting fully-nonlocal functional is exact for any collection of separated one-electron densities, and the PZ correction to the exact functional vanishes.

PZ SIC was an early attempt to improve approximate functionals via constraint satisfaction, and was proposed without much concern that satisfying an additional constraint or norm (exactness for all one-electron densities) might violate pre-existing constraints or norms (such as exactness for all uniform electron gases and 0% error for the exchange-correlation energies of neutral atoms in the limit of large atomic number [18]). A recent study [19] suggests that the SIC errors for these norms are not negligible.

When PZ SIC was proposed, LSDA was almost the only available density functional approximation for exchange and correlation. LSDA is not very accurate for the compact one- and two-electron hydrogen and helium atoms, and those errors are corrected by PZ SIC. The later GGA's and meta-GGA's also largely correct those errors, but they continue to make self-interaction errors for electron transfer and for the lobed one-electron densities that will be discussed in section 2.

The occupied orbital densities $n_{i\sigma}$ in Eq. (3) are not the squares of the occupied Kohn-Sham orbitals, which can be delocalized. To make SIC size-extensive, they should be the squares of unitarily-equivalent occupied localized orbitals. This was discussed in Ref. [17], but implemented to minimize the SIC energy by Pederson, Heaton, and Lin [20], enabling the first SIC calculations for molecules. The recently proposed Fermi-Löwdin orbitals [21], which are guaranteed to be localized, correspond to a particular set of real unitary transformations. Without such a restriction on the unitary transformation, delocalized orbitals that violate size-extensivity can be obtained when the underlying functional has a positive self-interaction correction from any localized orbital. This is unlikely for atoms and molecules in LSDA-SIC, but it can happen for PBE-SIC or SCAN-SIC, where the underlying functional is more accurate for compact one-electron densities.

In the absence of an external magnetic field or a spin-orbit interaction, the Kohn-Sham orbitals are usually chosen to be real (although complex Kohn-Sham orbitals may bring some benefits for orbital functionals [22]). Thus it was long assumed that the localized orbitals should also be real. Real orbitals that overlap one another must have nodes to ensure mutual orthogonality. Vydrov and Scuseria [23] showed that PZ SIC with real localized orbitals, applied to PBE, worsens the description of equilibrium (unstretched) bonds, and that conclusion is supported by later studies [24-27]. It has been suspected that noded localized orbitals limit the accuracy of PBE-SIC [28]. In 2011, Klüpfel, Klüpfel, and Jónsson [29] found that complex localized orbitals with nodeless orbital densities improve the accuracy of PBE-SIC over real ones. (While the real and imaginary parts of an orbital may have nodal surfaces, these surfaces can be different, leading to nodeless orbital densities [20,21].) Complex localized orbitals have also been found [25] to reduce but not eliminate PZ SIC symmetry-breaking in molecules. One argument [29] for their better performance is that introducing complex localized orbitals increases variational freedom and produces lower and more realistic PBE-SIC total energies. Refs. [25,29] found that complex orbitals yield lower energies than real ones for most systems. The problems that noded orbitals pose for semi-local functionals were also discussed in Refs. [28] and [30]. Hofmann, Klüpfel, Klüpfel, and Kümmel [30] say that "The smoother orbital densities without nodal planes are much closer to the realm where (semi-) local functionals are considered to be appropriate".

Here we will argue that semi-local approximations must fail for noded orbital densities for the same fundamental reason and in the same way that they must fail for stretched orbital densities. Perhaps the best-known failures of semi-local functionals occur for stretched orbital densities. In applications to ground states, the stretched orbitals are problematic for semi-local functionals, while the noded orbitals are problematic only for SIC functionals.

PZ SIC with real localized orbitals does indeed correct errors for stretched bonds, as shown in Fig. 1, but when applied to a semi-local functional it introduces other errors arising from the nodes of the orbital densities. The density-nodality error of PZ SIC is absent from uncorrected semi-local functionals applied to ground-state total densities, which are always nodeless for real systems. The most problematic ground states for the semi-local functionals are those of stretched radicals, such as $H_2^+$, $He_2^+$, transition states of chemical reactions, etc. Large errors in the total energies from semi-local approximations would also occur for strongly-stretched $H_2$ and $N_2$, if spin symmetry were not allowed to break [14,31].

It is well-known that the exact Kohn-Sham theory is exact for ground-state energies and densities, but not for most excited states of many-electron systems. The argument at the end of the next section shows that it is also exact for the excited states of one-electron systems.

## 2. Why semi-local approximations fail for lobed (e.g., stretched or noded) one-electron densities

A formally-exact expression for the exchange-correlation energy is [31,32]

$$E_{xc}[n_\uparrow, n_\downarrow] = \frac{1}{2} \int d\mathbf{r}\, n(\mathbf{r}) \int d\mathbf{r}'\, n_{xc}([n_\uparrow, n_\downarrow], \mathbf{r}, \mathbf{r}')/|\mathbf{r}' - \mathbf{r}|, \qquad (4)$$

i.e., a Coulomb interaction between the electron density at a position $\mathbf{r}$ and the density at $\mathbf{r}'$ of the exchange-correlation hole surrounding an electron at $\mathbf{r}$. Around a given electron, one electron is missing from the rest of the density, leading to the sum rule [31,32]

$$\int d\mathbf{r}'\, n_{xc}([n_\uparrow, n_\downarrow], \mathbf{r}, \mathbf{r}') = -1. \qquad (5)$$

The hole density is thus typically negative, and the more negative it is close to its electron the more negative the exchange-correlation energy will be. Although we do not usually know the hole density exactly, its known exact properties can be used to understand the power and limitations of approximate density functionals [31,32], and to provide an alternative and earlier construction [33] of the PBE GGA [4].

Semi-local approximations either explicitly [33] or implicitly [34] model the hole density as a function of $\mathbf{r}'$ that is localized (on a scale set by the local spatial variation of the electron density) around the electron position $\mathbf{r}$ and that satisfies the sum rule of Eq. (5). This model is reasonably correct in many cases. A functional like PBE or SCAN can be accurate both for the extended densities in solids and for the compact densities of atoms or of molecules at equilibrium geometries, while LSDA is less accurate, especially for the compact densities. However, for any one-electron density $n$ the exact hole density is $-n(\mathbf{r}')$, independent of $\mathbf{r}$, making Eq. (4) reduce to

$$E_{xc}^{exact}[n, 0] = -U[n] \qquad (6)$$

or minus the Hartree electrostatic energy of the one-electron density. Unless the one-electron density is reasonably compact, a semi-local approximation cannot mimic the full nonlocality of Eq. (6) (Fig.1). While LSDA cannot mimic Eq. (6), PBE and SCAN can do so for compact orbital densities (as in the small-$R$ part of Fig. 1). However, if a one-electron density is separated into two equal lobes (as in the large-$R$ part of Fig. 1, and in Fig. 2), then the exact hole density is shared equally between the lobes, while the semi-local hole density is concentrated on the lobe nearer to the electron, making the PBE or SCAN exchange-correlation energy too negative. In the limit of large separation between the lobes, the approximate hole is entirely on the lobe nearer the electron.

To illustrate this problem, Fig. 2 shows the electron density of the stretched one-electron ion $H_2^+$. The exact exchange-correlation hole density for any electron position $\mathbf{r}$ is $-n(\mathbf{r}')$ independent of $\mathbf{r}$, or Fig. 2 turned upside down. The LSDA hole density at $\mathbf{r}'$ for electron position $\mathbf{r}$ is a function of $|\mathbf{r}' - \mathbf{r}|$ that minimizes at the value $-n(\mathbf{r})$ when $\mathbf{r}' = \mathbf{r}$, then rises to zero over a length scale roughly proportional to $n(\mathbf{r})^{-1/3}$. This qualitative failure of LSDA is shared by GGA and meta-GGA for stretched $H_2^+$ and other stretched radicals.

From the very beginning of density-functional theory, it was known that the exact density functional for the exchange-correlation energy is nonlocal. For example, the exact exchange-correlation potential around an atom decays as $-1/r$ as $r \to \infty$, like a Coulomb potential. A more radical nonlocality, which does not decay to zero with increasing separation, was identified by Perdew, Parr, Levy, and Balduz 1982 [35], who were motivated in part by PZ SIC. This radical nonlocality is associated with non-integer average electron number on a separated subsystem. Stretched $H_2^+$, with half an electron (and half an exact exchange-correlation hole) around each separated proton, was identified as an example by Zhang and Yang 1998 [13], who found wrong dissociation limits with the BLYP GGA and B3LYP hybrid functionals, and made a quantitative model to explain this. It is clear from Ref. [13] that only a fully nonlocal functional like Hartree-Fock or PZ SIC can account for the energies of such systems. No semi-local approximation can be correct for such systems, where the error of a semi-local approximation grows gradually with increasing separation as in Fig. 3. The errors that semi-local approximations make for noded orbital densities reflect the incipient division of one electron into fragments of non-integer electron number.

GGA's and meta-GGA's can accurately describe compact one-electron densities such as the *1s* density in H or He$^+$. For example, the total energy of the hydrogen atom in hartree is -0.479 (LSDA), -0.500 (PBE or SCAN), or -0.500 (exact). It is well known that stretched orbitals, as in the stretched one-electron molecular ion H$_2^+$, are poorly described even by GGA's and meta-GGA's [13]. It is less well known (but see Ref. [36]) that noded orbitals, which are also lobed, encounter a similar self-interaction error. Figure 2 compares the densities of two different systems built up from one electron and two protons. The first is H$_2^+$ in its *1σ* even-parity ground state at the stretched bond length $R = 4$ *bohr*, with its density plotted along the bond axis. The second is the He$^+$ atomic ion in its *2p$_z$* excited state, with its density plotted along the z axis in Fig. 3. These systems are similarly lobed, and for both of them PBE or SCAN make large and remarkably similar relative errors (Table I). (The 2p$_z$ state of He$^+$ is introduced because its density is strongly noded, in much the same way that the H$_2^+$ density in Fig. 2 is strongly stretched. A noded density is the limit of a sequence of un-noded ground-state densities, as discussed at the end of this section.)

While a compact orbital density is clearly a single system with integer (1) electron number, stretching or noding can be regarded as early steps in the formation of separate systems with non-integer average electron number, for which semi-local approximations always fail dramatically.

Figure 3 shows that the PBE and SCAN errors almost go to zero as the bond length $R$ of H$_2^+$ is reduced to zero. Similarly, the large nodality errors for the *2p$_z$* orbital density are greatly reduced but not eliminated for typical real SIC localized orbitals of atoms and molecules (Table II). The *2p$_z$* orbital density is strongly noded, with a nodal plane through its center, but the real Fermi-Löwdin orbitals considered in Table II are more weakly noded, with nodal surfaces nearer their edges.

Since ground-state densities are typically nodeless, can we even define the exact ground-state exchange-correlation energy of a noded density? We can do so by regarding a noded one-electron density as the limit of a sequence of nodeless one-electron ground-state densities. For example, the *2p$_z$* orbital density in Fig. 1 (b) varies like $z^2$ near the nodal plane $z=0$, and can be obtained as the $a^2 \to 0$ limit of a sequence of one-electron densities that behave like $z^2 + a^2$ near z=0. A similar limit can be defined for any noded one-electron density, and the sequence of un-noded densities will all be *v*-representable ground-state densities by inversion of the one-electron Schrödinger equation for the positive square root of the density to find the potential *v(r)*. The integral of Eq. (2) then tells us that, for a noded as for a nodeless one-electron density *n*, Eq. (6) is valid. In other words, the exact ground-state exchange-correlation functional for spin-polarized one-electron ground-state densities, Eq, (6), applied to a sequence of nodeless spin-polarized one-electron ground-state densities that lie within its domain of definition, implies Eq. (6) even for noded spin-polarized one-electron densities. The exact exchange-correlation energy of a spin-polarized one-electron density is Eq. (6), which depends only on the amplitude and not on the phase of the one-electron wavefunction. Thus, as claimed in Ref. 17, the PZ correction on the extreme right of Eq. (3) does vanish when applied to the exact functional. Unfortunately, that does not imply that the correction will be acceptably small when applied to a semi-local functional in a situation for which that functional is accurate, as discussed in the next section.

From the discussion of the previous paragraph, the exact Kohn-Sham theory is exact for the ground *and* excited states of any one-electron density, although of course not for all excited states of many-electron systems, and PZ SIC is exact for all ground and excited states of one-electron systems.

### 3. Why complex localized orbitals work better than real ones in PZ SIC

GGA and meta-GGA functionals of the total density often work well, but not in regions where there are stretched orbitals. Our previous example was stretched $H_2^+$ [13], but the same effect occurs for the stretched bonds between different open-shell atoms [15] and in the transition states [16] of chemical reactions, where again the semi-local exchange-correlation energy is too negative, making the energy barriers too low.

A standard cure is the Perdew-Zunger (PZ) self-interaction correction (SIC) of Eq. (3). To work well, the PZ correction from each occupied localized orbital should be small (i.e., $E_{xc}^{approx}[n_{i\sigma}, 0]$ must nearly cancel $U[n_{i\sigma}]$), except when that orbital is stretched. But, as argued in the previous section, that cancellation will be less perfect when the SIC orbital density is noded (Tables I and II), and even less perfect as the orbital density becomes more strongly noded. As the orbital density becomes more noded, its semi-local exchange-correlation energy becomes more negative and its PZ self-interaction correction of Eq. (3) becomes more positive, even when the orbital density is unstretched. Noded SIC orbitals can be avoided, even when the canonical or generalized Kohn-Sham orbitals are real, by making the unitary transformation matrix and the resulting localized SIC orbitals complex, as in Refs. [25,29].

Tables III-V confirm this expectation for the atomization energies of molecules in the small representative set AE6 [37], for the barrier heights of chemical reactions in the small representative set BH6 [37], and for the 55 molecular formation energies of the G2-1 set [38]. For PBE and SCAN, the PZ self-interaction-corrected results are more accurate when complex SIC orbitals are used. These results for PBE-SIC are in agreement with those of earlier studies [23-27], which also noted that only the real localized orbitals are noded. But the explanation in the previous paragraph for the better results from complex orbitals is ours.

The SCAN and SCAN-SIC atomization energies of molecules (Tables III and V) illustrate the magnitude of the problem that remains to be solved. Without fitting to any bonded system, SCAN yields errors much smaller than those of LSDA or PBE. But SCAN-SIC yields errors that are bigger than those of SCAN by a factor of four or five for real SIC orbitals and three or four for complex SIC orbitals.

Tables III-V also agree with earlier studies [23-30] showing that PZ-SIC with real localized orbitals, applied to PBE (and in our study to SCAN), improves barriers but often worsens atomization energies. Use of complex orbitals solves only part of the problem by reducing but not eliminating the problematic lobed character of the orbital densities. (See Fig. 2 of Ref. [30].) The best SIC atomization energies in Table III are those of PBE-CSIC, as in Table I of Ref. [24]. Perhaps significantly, PBE-CSIC also yields the most accurate SIC total energies for large-$Z$ atoms.

PZ SIC imposes a new exact constraint (exactness for all one-electron densities) at the cost of other exact constraints or of appropriate norms such as exactness for the uniform electron gas and for large-$Z$ atoms [19]. This in turn results in errors that do not fully cancel out in the energy differences between free atoms and molecules. Needed is a generalization of PZ SIC that treats occupied localized orbitals in the PZ way only when they do not overlap with other occupied localized orbitals, but (unlike earlier attempts to scale down the self-interaction correction in many-electron regions [28]) retains the full Hartree part of the PZ SIC. Such a generalization is being attempted [39], to make SIC exact for all large-$Z$ atoms and other slowly-varying densities. While this seems to be achievable, restoring some of the other exact constraints satisfied by the nonempirical semi-local functionals may be harder to achieve with a fully nonlocal approximation.

## 4. Discussions and Conclusions

PZ SIC applied to a semi-local functional like the PBE GGA or the SCAN meta-GGA removes troublesome self-interaction errors that manifest most strongly for stretched bonds. But it introduces other errors, which manifest significantly in the atomization energies of molecules. Some of these are orbital-density nodality errors that can be removed by replacing real localized orbitals by complex ones (Tables III-V). Removing the nodes of the localized orbital densities reduces but does not eliminate their problematic lobed one-electron structure. Errors remain because the PZ-corrected functional for the total electron density violates (at least for real Fermi-Löwdin orbitals) [19] exact constraints or appropriate norms that are built into PBE and especially SCAN. Hopefully a *generalized* PZ SIC [39] will be able to recover some of these correct features of SCAN, thus providing a large self-interaction correction *only* where it is needed. SCAN, which is already self-correlation free, is a natural candidate for a generalized self-interaction correction. As a first step, we have here presented what may be the first results for the un-generalized PZ SIC applied to SCAN. (For a recent alternative approach, see Ref. [40].)

The PZ self-interaction correction does indeed vanish [17] when applied to the exact density functional for the exchange-correlation energy, but unfortunately that conclusion does not guarantee that, when applied to a good semi-local approximate functional, this correction will be acceptably small in situations where that semi-local functional is accurate. That is because the densities of occupied localized orbitals are typically more challenging to the semi-local functional than is the total electron density.

To make PZ SIC exact for any collection of separated one-electron densities, the right orbitals for the PZ SIC must span the space of occupied generalized Kohn-Sham or canonical orbitals. Using the often-delocalized generalized Kohn-Sham orbitals themselves would lead to a well-known size-extensivity problem [17]. We have here found a second reason not to use canonical orbitals: They can be highly noded (e.g., the *2p$_z$* orbital), with disastrous results for PZ SIC applied to semi-local functionals beyond LSDA (Table 1). A better choice is to use real localized orbitals that are more-weakly noded, i.e., noded near the edges but not near the center, such as the real Fermi-Löwdin orbitals [21]. Still better for the correction of semi-local functionals are complex localized orbitals with nodeless orbital densities [29]. Complex Fermi-Löwdin orbitals are possible [41], and also presumably guarantee the size-extensivity of SIC.

A good semi-local functional like PBE or SCAN will, by our analysis of section 2, produce a positive PZ self-interaction correction in Eq. (3) from any highly-stretched or strongly-noded localized orbital. As shown in Table VI, even the total self-interaction corrections to the exchange-correlation energies for atoms, using the weakly-noded real Fermi-Löwdin orbitals, tend to be positive for these functionals. Thus the strong nodality of some Kohn-Sham orbitals in atoms and small molecules drives the energy-minimizing localized orbitals away from the highly-noded Kohn-Sham orbitals and toward the more weakly-noded Fermi-Löwdin orbitals. But, even so, the PBE-SIC and SCAN-SIC exchange-correlation energies with real SIC orbitals are not negative enough, and are much less accurate than those of SCAN (Table VI).

PZ SIC reduces atomization energies of molecules (often by too much) and raises energy barriers to chemical reactions. We can now propose a tentative explanation for this, although the explanation is restricted to the case of real localized SIC orbitals: In the ground state, real localized SIC orbitals that overlap must be noded to achieve orthogonality. As we pass from separated atoms to the molecules they form, or from separated molecules to the transition states they form, the valence orbitals acquire more orbitals with which they overlap, and hence become more noded. As the valence orbitals become more noded, their PZ self-interaction correction of Eq. (3) becomes more positive or less negative. Thus PZ SIC raises the energy of a molecule relative to the energy of its separated atoms, and raises the energy of a transition state relative to its separated molecules. PZ SIC also raises the energy of the transition state relative to its separated molecules through bond stretching.

The correlation energy as defined in the exact Kohn-Sham theory is numerically close to its quantum-chemistry definition [42], the deviation of the total energy from the Hartree-Fock value. In this sense, the transition states of chemical reactions are more strongly-correlated than are the ground states of molecules in equilibrium, since the Hartree-Fock energy barriers to chemical reactions are seriously too high. The barriers from semi-local functionals are seriously too low, suggesting that in this sense they "over-correlate" for stretched radicals, but the barriers are rather accurate when a self-interaction correction is applied. This and other results (e.g., Refs. [9,43]) suggest that properly-self-interaction-corrected semi-local functionals might usefully describe the energetics of strongly-correlated systems. By properly-self-interaction-corrected, we mean exact for all one-electron densities and reasonably accurate even for fractional electron number, as in PZ SIC, but roughly as accurate as SCAN for atomization energies and equilibrium bond lengths (an accuracy that arises from the satisfaction of exact constraints and appropriate norms, with no fitting to bonded systems). The SCAN and SCAN-SIC results in Tables III, V, and VI show how far we still are from this goal. Hybrid functionals [43], that mix an empirical fraction of exact exchange with a complementary fraction of semi-local exchange, go only part of the way toward this goal by reducing but not eliminating the self-interaction error.

It is not only atomization energies but also equilibrium bond lengths that are worsened by PZ SIC. For small molecules, LSDA-SIC bond lengths are shorter and less accurate than those of LSDA, and PBE-SIC bond lengths are shorter and less accurate than those of PBE, whether real [28] or complex [44] SIC orbitals are used. In addition, PZ SIC breaks molecular symmetries [25]. We plan to return to these problems in future work.

A by-product of this work is the observation that the exact Kohn-Sham theory, with the standard exact ground-state exchange-correlation energy functional, is exact for all excited states of one-electron systems, as well as for all ground-states, and that the Perdew-Zunger self-interaction correction [17] (PZ SIC) perfectly imitates the exact Kohn-Sham theory in this special case of one-electron densities. For a many-electron excited state, of course, there is no exact Kohn-Sham theory, but removing the self-interaction error from an approximate functional could improve the approximate description of an excited state.

**Supplementary Material**

The supplementary material presents details of the ERKALE and PySCF/PyFLOSIC codes, and detailed results for the AE6, BH6, and G2-1 molecular test sets.

**Acknowledgments:** The work of many of us (BS, KAJ, JEP, KT, HM, SA, AR, YY, TB, RRZ, and JPP) was supported by the US Department of Energy, Office of Science, Basic Energy Sciences, under Award No. DE-SC0018331 as part of the Computational Chemical Sciences Program. The work of CS was supported by the Center for Complex Materials from First Principles, an Energy Frontier Research Center funded by the US Department of Energy, Office of Science, Basic Energy Sciences, under Award No. DE-SC0012575. The work of PB and KW was supported by the US National Science Foundation under Grant No. DMR-1607868. The work of JK was supported by the German Research Foundation (DFG) within CRC920-A04. SL acknowledges support from the Academy of Finland through project number 311149. NKN acknowledges support by the National Science Foundation under Grant No. DMR-1553022. Many of us acknowledge stimulating discussions with Mark R. Pederson. JPP designed the work and wrote the first draft. The other authors contributed calculations, figures, tables, references, discussions, and revisions. This research includes calculations carried out on Temple University's HPC resources, and thus was supported in part by the National Science Foundation through major research instrumentation grant number 1625061 and by the Army Research Laboratory under contract number W911NF-16-2-0189.

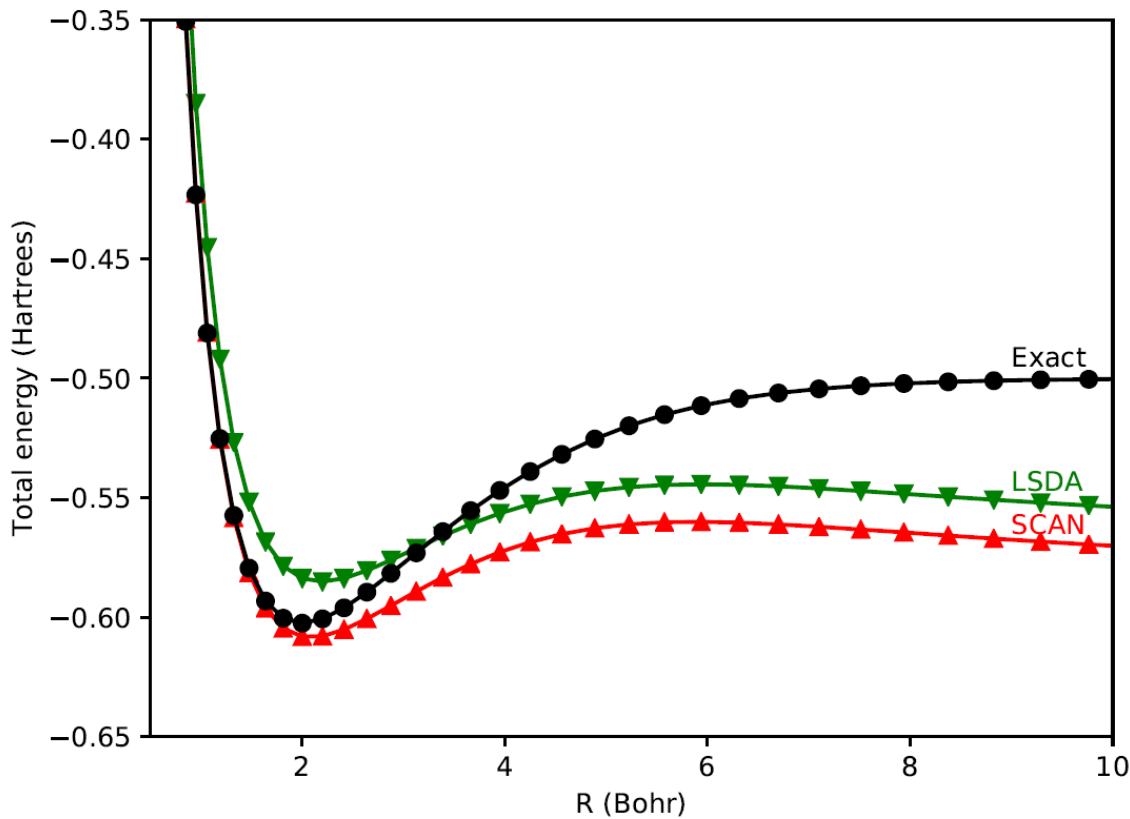

Fig. 1. Binding energy curves: exact (or equivalently SIC), LSDA, and SCAN total energies as functions of bond length $R$, for the one-electron molecular ion $H_2^+$, calculated from self-consistent densities with the PySCF code [45,52] and the cc-pVQZ [46] Gaussian basis set. Note that SCAN (as well as PBE, not shown) is accurate for the compact one-electron densities of compressed or equilibrium bond length, but makes the energy increasingly too negative as the bond is stretched. (Note further that, beyond one- and two-electron densities, SCAN and SIC make greater demands on the mesh for the integral of Eq. (1) than LSDA or PBE do. For example, in the FLOSIC [49] code, the volume per mesh point is 3.7 times smaller for SCAN than for LSDA or PBE.)

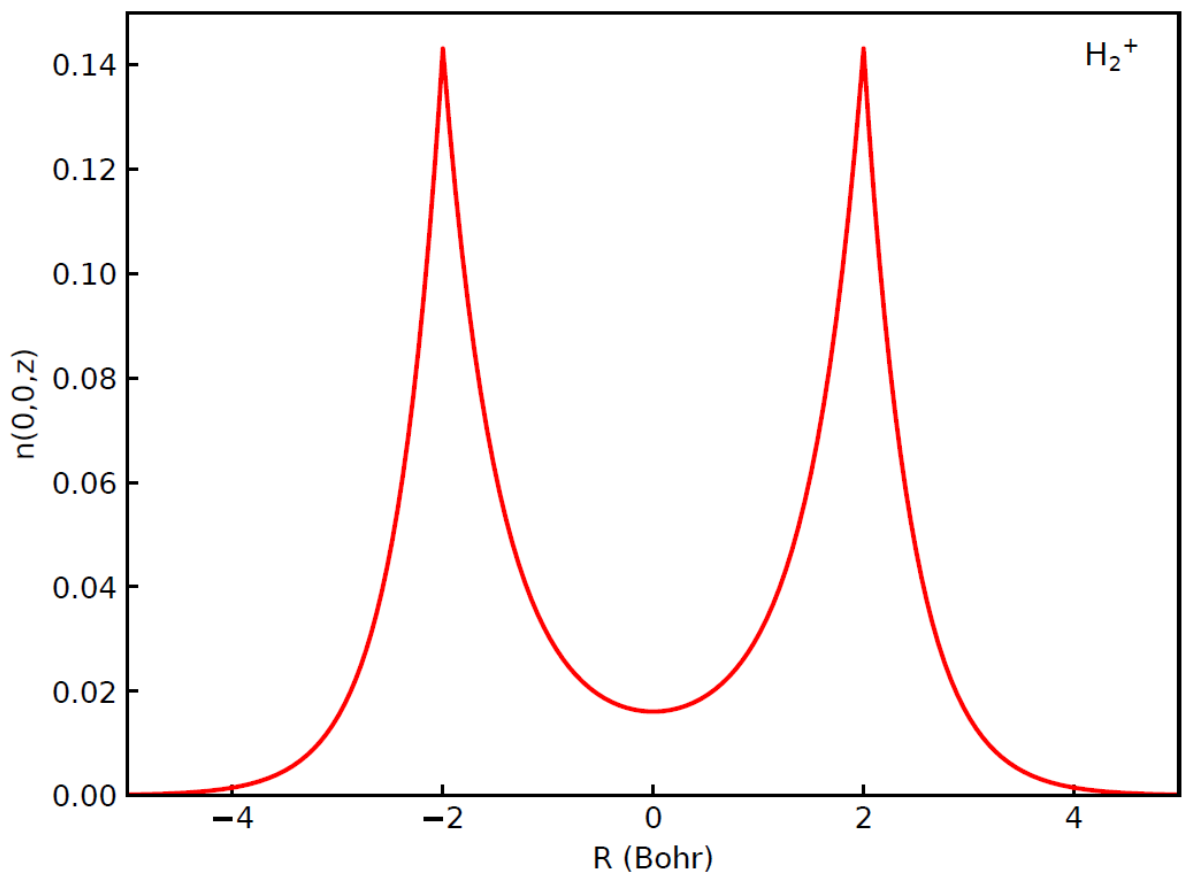

Fig. 2. (a) Exact (Hartree-Fock or PZ SIC) non-relativistic electron density (in atomic units) of the one-electron molecular ion $H_2^+$ in its ground state at stretched bond length *R = 4 Bohr*, computed in the complete basis-set limit using the code HelFEM [47,48]. A similar density with slightly lower and more rounded nuclear cusps was obtained with the FLOSIC [49] all-electron Gaussian-type-orbital code using the NRLMOL default basis set [50].

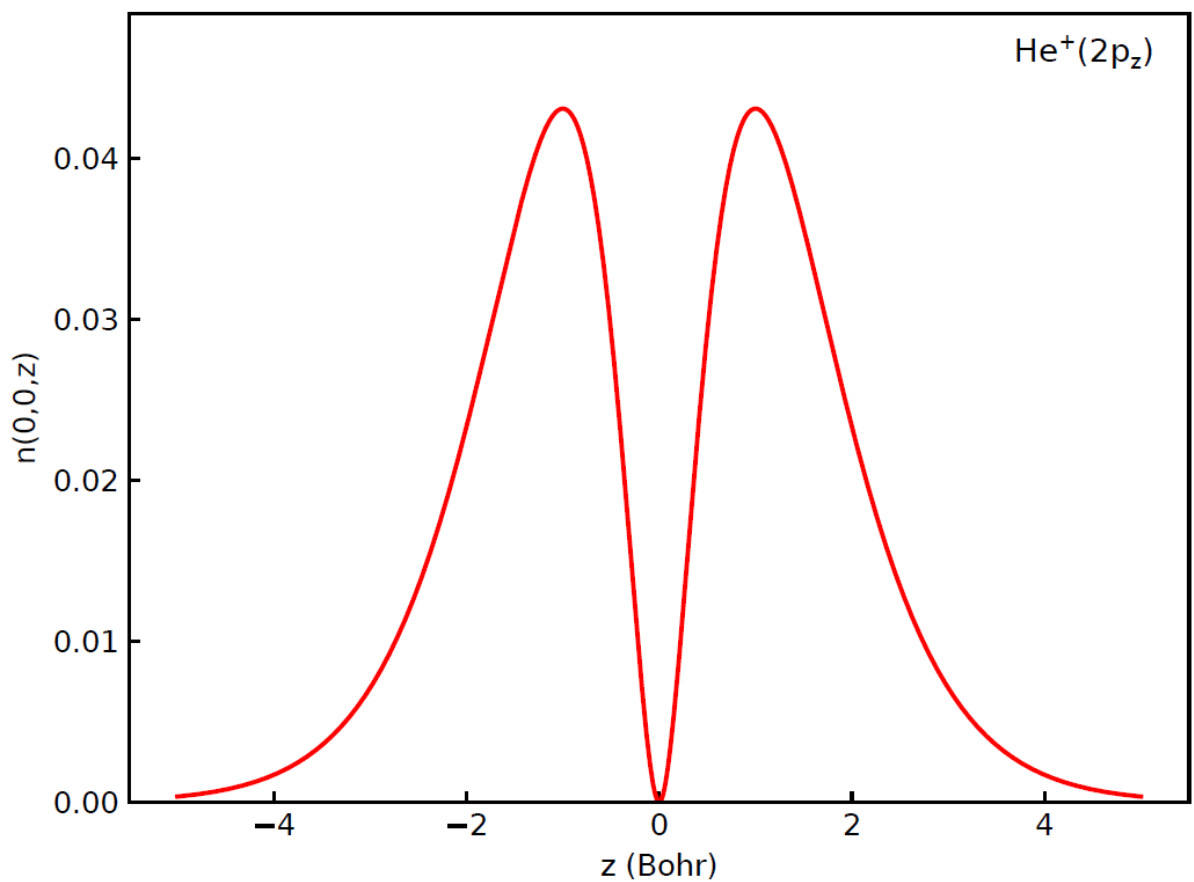

Fig. 2. (b) Electron density (in atomic units) of He$^+$ in its noded *2p$_z$* excited state, computed using the exact non-relativistic analytic expression. The one-electron density of Fig. 2 (a) is stretched, and that of Fig. 2 (b) is noded, but both are similarly lobed.

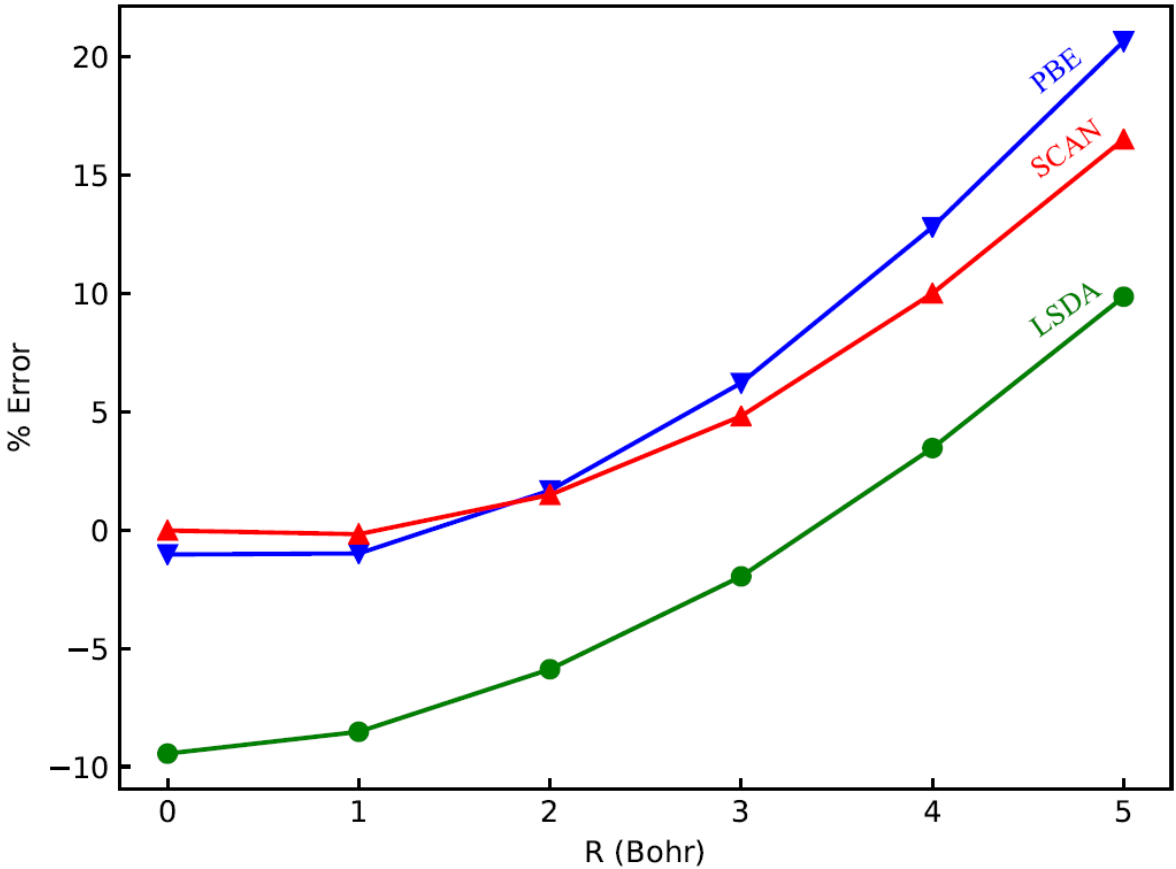

Fig. 3. Relative error of LSDA, PBE, and SCAN for $E_{xc}$ of $H_2^+$, using exact (i.e, SIC) electron densities from the FLOSIC code [49], as functions of bond length $R$. Note that at $R=0$ the exchange-correlation energy is that of $He^+ 1s$. The equilibrium bond length is around *2 Bohr* (Fig. 1). The results for the non-zero bond lengths of $H_2^+$ correct erroneous results from Ref. [36]. For PBE and SCAN, the self-interaction error grows as the bond is stretched, but it grows more slowly in SCAN.

Table I. Relative errors (%) of LSDA, PBE, and SCAN for $E_{xc}$ of $H_2^+$ at $R = 4$ *Bohr* and $He^+$ $2p_z$, both using exact densities. The $H_2^+$ values from PySCF [45,52] using the cc-pVQZ [46] Gaussian basis set correct erroneous values from Ref. [36]. Exact and approximate exchange (but not correlation) energies of the $He^+$ canonical orbitals scale up in proportion to the nuclear charge Z. Note the similarities in the numbers between the stretched $H_2^+$ and the noded $He^+$ $2p_z$ densities. Note further that SCAN is better than PBE for these highly stretched or strongly noded one-electron densities. The bottom row shows the exact $E_{xc}$ of Eq. (6), in Hartrees.

| Approx. | Stretched $H_2^+$ | $He^+$ $2p_z$ |
|---|---|---|
| LSDA | 3.6% | 3.4% |
| PBE | 12.9% | 12.8% |
| SCAN | 10.1% | 8.8% |
| $E_{xc}$ exact | -0.2285 | -0.1957 |

Table II. Relative errors of LSDA, PBE, and SCAN for $E_{xc}$ of individual real Fermi-Löwdin orbitals in the Ne atom and the CH$_4$ molecule, calculated here using the corresponding SIC orbital densities from the FLOSIC [49] code. The exact entry $-U[n_{i\sigma}]$ is displayed for the SCAN-SIC orbital density. For the Ne atom, the core orbital is *1s* and the valence orbitals are rotationally-equivalent *sp³* hybrids. The purpose of this table is not to compare atoms with molecules, but to show that the relative errors of PBE and SCAN for the valence orbitals are much smaller here than those in Table I, because the real Fermi-Löwdin orbital densities are less severely noded than is the $2p_z$ orbital density. (As usual, PBE and SCAN are much more accurate than LSDA for the compact *1s* core orbital densities.) The bottom row shows the exact $E_{xc}$ of Eq. (6), in Hartrees. For CH$_4$, we have employed the QCISD/MG3 equilibrium geometry (https://comp.chem.umn/db.index-html).

| Approx. | Ne core | Ne valence | CH$_4$ core | CH$_4$ valence |
|---|---|---|---|---|
| LSDA | -12.5% | -6.3% | -11.6% | -6.4% |
| PBE | -1.8% | 3.9% | -1.5% | 1.2% |
| SCAN | 0.2% | 3.2% | 0.2% | 1.0% |
| Exact | -3.065 | -0.589 | -1.781 | -0.345 |

Table III. Mean error (ME) and mean absolute error (MAE) (kcal/mole) for LSDA, PBE, SCAN, LSDA-RSIC, PBE-RSIC, SCAN-RSIC, LSDA-CSIC, PBE-CSIC, and SCAN-CSIC for the molecular atomization energies of the small representative AE6 set. RSIC is SIC with real localized orbitals, and CSIC is SIC with complex localized orbitals. The AE6 [37] set comprises the six molecules $SiH_4$, $S_2$, SiO, $C_3H_4$ (propyne), HCOCOH (glyoxal), and $C_4H_8$ (cyclobutane). All densities are self-consistent. We have used the Gaussian-type-orbital ERKALE code of Lehtola et al. [51,52] with the cc-pVQZ [46] basis set, using fixed nuclear geometries at the QCISD/MG3 level (see https://comp.chem.umn.edu/db/index.html). A Lebedev-Lakov grid of 50 radial and 194 angular points delivered good agreement with reference values [53]. See the Supplemental Material for further details. (1 Hartree = 627.5 kcal/mole = 27.21 eV.) Note that here PZ SIC worsens the atomization energies of SCAN, but less severely in CSIC than in RSIC.

|           | AE6   |      |
|-----------|-------|------|
| approx    | ME    | MAE  |
| LSDA      | 77.3  | 77.3 |
| LSDA-RSIC | 57.1  | 60.0 |
| LSDA-CSIC | 62.6  | 62.6 |
| PBE       | 12.5  | 15.6 |
| PBE-RSIC  | -14.4 | 17.8 |
| PBE-CSIC  | -8.6  | 10.0 |
| SCAN      | -2.2  | 4.4  |
| SCAN-RSIC | -23.0 | 24.3 |
| SCAN-CSIC | -16.9 | 17.1 |

Table IV. Same as Table III, but for the barrier heights of the small representative BH6 [37] set. There are forward and backward barriers for each of three reactions: (1) $OH+CH_4 \rightarrow CH_3+H_2O$. (2) $OH+H \rightarrow O+H_2$, and (3) $H+H_2S \rightarrow HS+H_2$. Note that PZ SIC improves the barrier heights for all the semi-local functionals.

|           | BH6   |      |
|-----------|-------|------|
| Approx.   | ME    | MAE  |
| LSDA      | -18.1 | 18.1 |
| LSDA-RSIC | -5.1  | 5.1  |
| LSDA-CSIC | -4.1  | 4.1  |
| PBE       | -9.6  | 9.6  |
| PBE-RSIC  | -0.2  | 4.1  |
| PBE-CSIC  | 2.2   | 2.2  |
| SCAN      | -7.9  | 7.9  |
| SCAN-RSIC | -1.2  | 2.7  |
| SCAN-CSIC | -0.7  | 2.3  |

Table V. Same as Table III, but for the 55 molecular formation energies of the G2-1 data set [38], using SCAN, SCAN-RSIC, and SCAN-CSIC. The formation energies are constructed in such a way that their MAE's are essentially those of the molecular atomization energies, and their ME's are essentially minus those of the atomization energies. We have used B3LYP geometries with zero-point expansion. Note that the FLOSIC method implemented in the PySCF code [45,52] (PyFLOSIC [54]) agrees within 0.6 kcal/mole with these RSIC results from the ERKALE code [51.52]. (See the Supplemental Material for details). For LSDA, LSDA-RSIC, PBE, and PBE-RSIC, see Ref. [27].

|  | G2-1 | |
| --- | --- | --- |
| Approx. | ME | MAE |
| SCAN | 1.73 | 4.08 |
| SCAN-RSIC | 14.2 | 16.0 |
| SCAN-CSIC | 8.9 | 10.2 |

Table VI. Total PZ self-interaction corrections to the exchange-correlation energies of rare-gas atoms, in Hartrees. Shown is the orbital-density-dependent correction of Eq. (3) to the exchange-correlation energy in an energy-minimized SIC calculation, with real (RSIC) or complex (CSIC) orbitals, using the ERKALE code [51,52] and the standard NRLMOL basis set [50]. Note that the CSIC values are lower than the RSIC values, as expected, and that the difference increases from LSDA to PBE to SCAN. The CSIC correction to PBE is relatively small, consistent with Ref. [29]. To set a scale, the last rows show the accurate total exchange-correlation energy from a self-consistent SCAN calculation, and a nearly-exact exchange-correlation energy (exchange from Ref. [55] and correlation from Ref. [18]).

| Functional | Ne | Ar | Kr | Xe |
|---|---|---|---|---|
| LSDA-RSIC | -1.06 | -2.61 | -7.55 | -13.74 |
| LSDA-CSIC | -1.11 | -2.75 | -8.02 | -14.57 |
| PBE-RSIC | 0.08 | 0.26 | 1.28 | 2.99 |
| PBE-CSIC | -0.04 | -0.08 | -0.01 | 0.56 |
| SCAN-RSIC | 0.16 | 0.46 | 1.76 | 3.80 |
| SCAN-CSIC | -0.20 | -0.65 | -3.54 | -8.08 |
| Total $E_{xc}$ (SCAN) | -12.45 | -30.99 | -95.79 | -181.99 |
| Total $E_{xc}$ (exact) | -12.50 | -30.91 | -95.74 | -182.20 |